# Evaluation of Reusability in Aspect Oriented Software using Inheritance Metrics

Vinobha A[1], Senthil Velan S[2], Chitra Babu[3]
[1]PG Scholar
[1,2,3]Department of Computer Science and Engineering, SSN College of Engineering, Chennai, India
[1]vinobha1217@cse.ssn.edu.in, [2]senthilvelan@ssn.edu.in, [3]chitra@ssn.edu.in

*Abstract*—Aspect-Oriented Software Development *(AOSD)* is a promising methodology for efficiently capturing the cross-cutting functionalities (concerns) as independent units called aspects. Inheritance of classes and aspects play a vital role in defining the units of encapsulation. Hence, it is essential to quantitatively capture the impact of inheritance in *AOSD* using design level metrics and to infer on the higher level quality attribute, reusability. An application to automate the processes of a typical University has been developed in order to study the effect of using inheritance over the versions of an aspectized *AO* application. A set of metrics to capture the manifestations of inheritance is proposed for measurement. An automated tool named as Aspect Oriented Software Reusability Measurement *AOSRM* is also designed and developed to calculate the values of the proposed metrics. Based on the obtained metric values for Java and AspectJ versions of the case study application, inheritance in AspectJ versions showed a positive impact on reusability of software.

## I. INTRODUCTION

Aspect Oriented Software Development *(AOSD)* methodology lets designers as well as programmers to neatly isolate and encapsulate the cross-cutting concerns into independent modular units [1]. This increase in modularity enables software designers to develop the components of software keeping in mind to have a higher degree of reusability.

Inheritance is one of the powerful notions in modular software development which can be used to extend or redefine the functionalities encapsulated in Object Oriented *(OO)* software. A good number of software developed using *OO* methodology uses inheritance in various forms for the efficient utilization of the encapsulated entities. Though inheritance is a fundamental property for extension, there are several complex issues and consequences in applying the concept, which is widely debated in the literature. Hence, the impact of inheritance in any methodology needs be thoroughly studied. The concept of inheritance in *AOSD* methodology introduces additional set of elements such as abstract and concrete aspects. These abstractions are very similar to the base and derived classes defined in *OO* based software development.

The consequences of applying the concept of inheritance in *AOSD* methodology need to be explored further. This will enable the software designers to understand its impact towards the higher level quality attribute namely reusability. In order to do so, the *AO* software needs to be quantitatively evaluated with a set of metrics that captures the manifestation of inheritance in its constructs [3]. A method in which a part of software design can be reused by adding additional functionalities onto the existing one with little or no modification is called as reusability.

The reusability of a software can be enhanced by applying one of the crucial design property namely inheritance [6]. Inheritance can be effectively used to extend or redefine the existing functionalities with clearly defined rules. Implementation inheritance is a mechanism whereby a subclass reuses the code in a base class to reuse the functionality. Either the subclass can retain all the operations of the base class or the operations can be redefined to add new functionalities during the evolution of software. When a class is reused in order to model the changes in requirement there will be a reduction in development time and consequently the testing time of software.

The remainder of this paper is organized as follows. Section II discusses about the existing work on inheritance of artifacts. Section III describes the motivation behind the work done. Section IV and Section V present the methods of inheritance and the proposed class aspect inheritance model for *AO* software. Section VI describes the approach to measure inheritance using metrics. Section VII describes the various artifacts, manifestations of inheritance and proposed metrics. Section VIII explains the various functionalities involved in the case study application. Section IX explains the architecture of the *AOSRM* tool that is developed to automate the evaluation of the proposed metrics. Section X presents a discussion on the obtained values of metrics across the versions of the case study application. Section XI concludes with possible directions for enhancement.

## II. EXISTING WORK

A series of experiments has been done to test the effect of inheritance on maintainability of *OO* software. Daly [6] found that, *OO* software can be developed 20 percent quicker when inheritance is used for software development. It is also found that 2 out of 3 subjects i.e., (lecturer, staff and professor) performed faster when inheritance is applied to develop *OO* software. In this work, three levels in the depth of inheritance have been focused only in *OO* software. Hence there is a need to analyze the impact of designing *AO* software with three levels or more than three levels of inheritance.

Hanenburg et al. [2] stated that advices in aspects are not treated on par with methods of classes. Since advices of aspects do not have a name, there is no way to decide whether the developer wants to redefine the same advice or would like to introduce another advice. The advices cannot be extended further, since all advices are not named and identified only through the respective pointcut. Hence, it is required to define a method inside advice to override the advice in the abstract aspect. This work focused only on the advices of aspects to explain the cases of inheritance. Hence, there is a clear need to focus on other constructs in *AO* software.

Bansiya et al. [8] proposed a hierarchical quality model to capture the design properties from the available *OO* metrics. The captured design properties are further used to infer on higher level quality attributes. This work was not extended to measure the effect of inheritance on *AO* software.

Jarallah AlGhamdi et al. [7] proposed a tool for measuring inheritance coupling in *OO* software. This tool has three components, namely, parsing engine, central query repository and query engine. It collects information about the factors that affect the coupling due to inheritance coupling with the help of measurements from *OO* software through its parser component. This automated tool had been used to capture only the design property namely coupling and further inference on the reusability has not been attempted.

P. Gulia [4] proposed two new inheritance metrics, *CCDIT* (Class Complexity due to Depth of Inheritance Tree) and *CCNOC* (Class Complexity due to Number of Children) to measure the complexity of methods in classes. These metrics were proposed to measure inheritance only for *OO* software.

III. MOTIVATION

Inheritance of classes and aspects plays a vital role in defining the units of encapsulation. Even though inheritance is a basic notion to achieve reusability, if the levels of hierarchy are increased too far it then reflects on the complexity of software. Therefore, the consequence of applying the concept of inheritance in *AOSD* needs to be explored through empirical evaluation. This enables the designer to understand the impact towards higher level quality attribute of reusability. Therefore, metrics are required to quantitatively measure the effect of inheritance in *AOSD* methodology.

IV. INHERITANCE IN *AO* SOFTWARE

Methods of achieving inheritance in *AO* software is shown in Fig. 1. Aspect inheritance introduces two additional set of elements such as abstract and concrete aspects [2]. Abstract aspect can act as the base aspect as like base class in *OO* where we can define the functionalities and concrete aspect is like derived class in *OO*. It inherits the functionalities which were given in abstract aspect to reuse it or to override it. As shown in Fig. 1, a number of abstract and concrete aspects can be inherited from any given abstract aspect. A concrete aspect cannot be further extended. Abstract and concrete aspects can inherit from a class or an interface, but vice-versa is not possible.

V. CLASS ASPECT *(CA)* INHERITANCE MODEL

The concept of Class Aspect Inheritance model is shown in the Fig. 2. This model shows the inheritance depth possible through both the levels of classes and aspects. Derived class C can inherit from base classes 1 ... P. Similarly, concrete aspects 1 ... M can be woven in the derived class C. Each concrete aspect can in turn be extended from multiple levels of abstract aspects, 1 ... O or 1 ... N. Through the *CA* inheritance model, the following effects which are listed below can be measured.

- The effect of having multiple concrete and abstract aspects whose pointcuts advice the join points present only in the derived class C.
- The effect of having multiple concrete and abstract aspects whose pointcuts advice the join points present only in the base class of the derived class C.

In order to capture these effects of inheritance in both classes and aspects, a set of metrics has been proposed and defined using proper mathematical notations.

A. *Existing metrics for measuring inheritance*

In Aspect oriented software development, it is necessary to measure the inheritance of aspects in order to understand the reuse of aspects effectively. There are some metrics for measuring the inheritance of objects in Object-Oriented software [3]. The metrics proposed by Chidamber and Kemerer contains two metrics that quantifies the inheritance of objects, namely, *DIT* and *NOC*.

*1) DIT (Depth of Inheritance Tree):* The metric *DIT* can be defined as the maximum distance from the node to the root in a tree. It is also a measure of the number of ancestor classes that are affected by the node. It is difficult to predict the behavior of a class through its modeled methods, if the class is found to be in the lower level of the class hierarchy with more inherited methods [4]. It can also be said that deeper trees exhibit the greater design complexity, since more methods and classes are involved.

*2) NOC (Number of Children):* Number of Children measures the subclasses of an intermediate class in the class hierarchy tree. It actually measures the impact of a change in a class on its immediate subclasses.

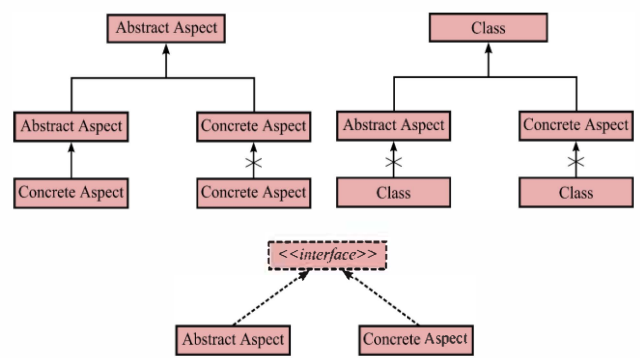

Fig. 1 Methods of Inheritance for Artifacts in *AO* software

## VI. PROPOSED WORK

Since, reusability cannot be directly measured from the design of software, there is a need to develop a quality model to quantitatively assess the reuse of components and indirectly infer on the higher level quality attributes. In both *OO* and *AO* reusability is achieved through inheritance of classes and aspects. These manifestations of inheritance need to be quantitatively assessed using well defined software metrics. Based upon the measured values, the impact of *AO* design on inheritance can be inferred and further on the higher level quality attribute, reusability. The effect of inheritance during the evolution of both *AO* and *OO* software can be analyzed to deduce the strengths of applying *AOSD* methodology.

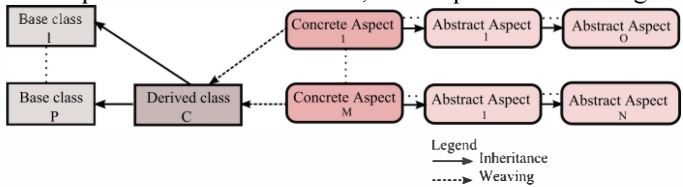

Fig. 2 Class Aspect Inheritance Model

## VII. *AO* REUSE EVALUATION MODEL

Set of manifestations for inheritance has been defined and its occurrence in *AO* software has been identified. Metrics are defined to quantitatively measure the manifestations of inheritance in *AO* software, infer on the effect of inheritance and further on the higher level quality attribute, reusability. The defined manifestation and proposed metrics are shown in the Fig. 3.

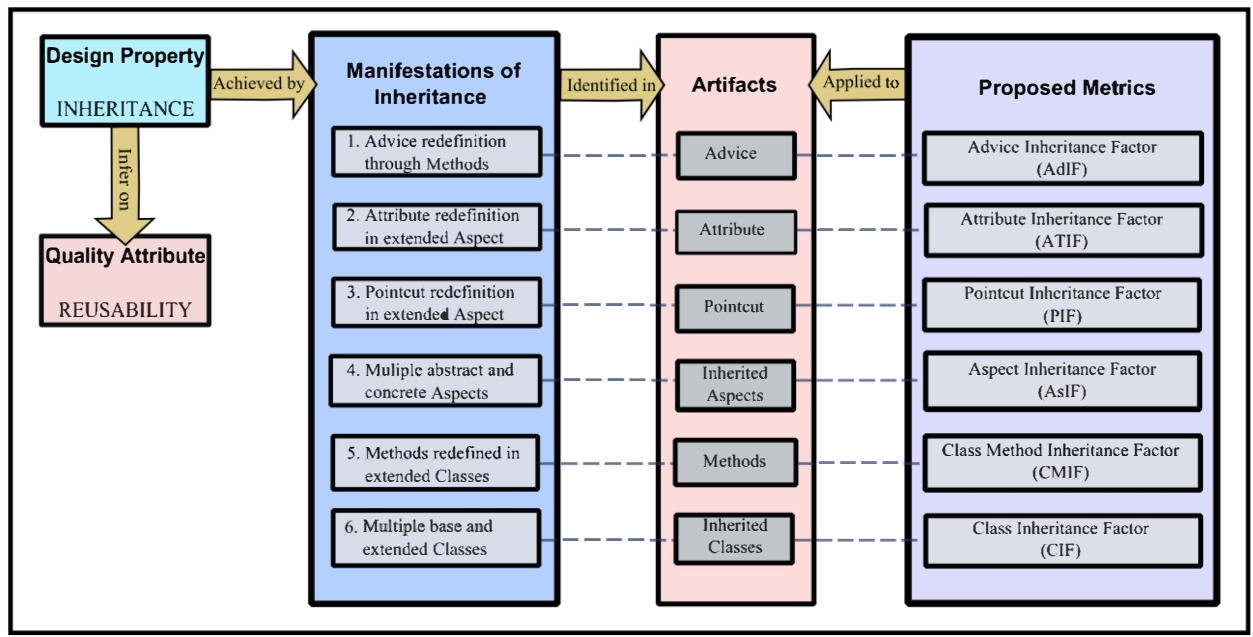

Fig. 3 *AO* Reuse Evaluation Model

### A. Artifacts

Artifacts are the modeling elements considered in which the set of manifestations are realized. The artifacts identified include both *AO* and *OO* modeling elements.

*1. Advice:* Advice is the code segment that is executed at each join point chosen through a pointcut [9]. There are three kinds of advice: *before()*, *around()*, and *after()* advice. As their names suggest, *before()* advice is executed before the join point, *around()* advice is executed by optionally bypassing the join point and *after()* advice is executed just after the execution of the join point.

*2. Attributes:* Attributes enable programmers to write flexible programs. A developer uses variables to represent data instead of entering data directly into the program. While executing the program, variables hold the real data. This only enables the same program to process different sets of data.

*3. Pointcut:* A pointcut is a construct to represent a set of join points. Whenever the program execution reaches one of the join points defined in the pointcut, a piece of code known as advice associated with the pointcut is executed [9]. This describes where and when additional code should be executed to an existing behavior. It also permits the weaving of aspects to the existing software.

*4. Methods:* Methods are constructs inside an object with which the functionalities of the object are defined. Methods of objects are usually invoked through a clearly defined set of interfaces i.e., methods implement the set of behavior for the objects. Each method has a name, input parameters and a return type.

*5. Classes:* Objects can be created inside a class. Constructor is mainly used to create objects. The objects which is created by the constructor is said to be an instance of the class. An instance variable is the member variables which is specific to the object.

### B. Manifestations of inheritance

MANIFESTATION 1. *Advice redefinition through methods*.

Advice is one of the important artifacts defined in *AO* software. A pointcut defines the set of join points in *AO* software and contains the signatures of the join points. The respective advice defined through the pointcut is executed when a join point is reached during program execution. In order to reuse a function defined in an abstract aspect, the corresponding advice need to be overridden. In *AO*, advice redefinition is not possible since advices do not have a name. However, the method invoked by an advice of the base aspect can be redefined through inheritance of aspects. This method of redefinition is required since there is a possibility for a change in the behavior of method over time. Hence, to override the code implemented in an advice, methods can be called in the advice and the methods can define the behavior.

For example, in an Online shopping software, objects can be *shoppingCart, customer and product*. Their behaviors are defined by *placeOrder(), makePayment()* and *offerDiscount()* respectively. There are possibilities that *makePayment()* method can change over time. During festival time, some special reduced prize offers are provided and hence, this functionality can change over time. In such cases, this manifestation can be applied to extend or redefine the functionality.

MANIFESTATION 2. *Attribute redefinition in extended aspect*.

Attribute is another artifact defined in both *OO* and *AO* versions of software. Attribute defines the type and state of the object and hence, when the type or state of the object changes, there is a need to modify or redefine the attribute. Attributes defined in base aspect can be modified or redefined in the extended aspect.

For example, in a shopping cart application, the system needs to perform round off when the price of the particular product is 99.9 and has to store in the same prize attribute. Hence, if the type of price is defined as float then it has to be refined as double in the extended aspect.

MANIFESTATION 3. *Pointcut redefinition in extended aspect*.

Pointcuts in *AO* software are similar to methods in Java programming language. A pointcut can be named or unnamed, just as a class can be named or anonymous in *OO* software. Pointcut defined in base aspect can be redefined in the extended aspect. Using pointcut redefinition, a particular feature of the base class can be reused in the extended class. It only enables the programmer to adapt the class based on the needs of extension.

For example, there are several products in Online shopping system such as CD player and DVD player. Behaviors like *placeOrder()* and *makePayment()* is common for both the products. If a DVD player inherits the CD player then DVD player is compatible for adding some extra functionality like offering discount under certain conditions. In such cases pointcut redefinition is done in *AO* software.

MANIFESTATION 4. *Multiple concrete and abstract aspects*.

AspectJ compiler allows only the inheritance of abstract aspects and not from a concrete aspect. In addition, if the derived aspect attempts to override a concrete pointcut in the base aspect, a compile-time error will occur. In a concrete aspect, the use of the *extends* keyword tells the AspectJ compiler that the concrete aspect needs to define each abstract method and pointcut in the abstract aspect.

For example, persistence is one of the important cross-cutting concerns implemented in University Automation System *(UAS)* case study. In the case study, database persistence can be implemented in an abstract aspect, since it performs the common database connection. Whereas, SQL translate and all the SQL operations such as *insert(), update()* and *delete()* can be implemented in the concrete aspect. The concrete aspect can extend the abstract aspect whenever there is a need for database connectivity.

C. *Proposed metrics*

The quantification of the manifestations discussed in the previous section is done by a set of metrics proposed in this section.

1. *Advice Inheritance Factor* (*AdIF*): Advice Inheritance Factor is the fraction of the sum of number of redefined advices in all aspects over the sum of number of available advices in all aspects defined in a version of *AO* software.

$$AdIF = \frac{\sum_{i=1}^{n} A_r(A_i)}{\sum_{i=1}^{n} A_a(A_i)} \quad (1)$$

Where,

$Ad_r(A_i)$ is the number of redefined advices defined in aspect $A_i$,

$Ad_a(A_i)$ is the number of available advices defined in aspect $A_i$ and

$n$ is the number of aspects in the version of *AO* software.

2. *Pointcut Inheritance Factor* (*PIF*): Pointcut Inheritance Factor is the fraction of the sum of number of redefined pointcuts in all aspects over the sum of number of available pointcuts in all aspects defined in a version of *AO* software.

$$PIF = \frac{\sum_{i=1}^{n} P_r(A_i)}{\sum_{i=1}^{n} P_a(A_i)} \quad (2)$$

Where,

$P_r(A_i)$ is the number of redefined pointcuts defined in aspect $A_i$,

$P_a(A_i)$ is the number of available pointcuts defined in aspect $A_i$ and
$n$ is the number of aspects in the version of *AO* software.

3. *Attribute Inheritance Factor* (*AttIF*): Attribute Inheritance Factor [5] is the fraction of the sum of number of redefined attributes in all aspects over the sum of number of available attributes in all aspects defined in a version of *AO* software.

$$AttIF = \frac{\sum_{i=1}^{n} Att_r(A_i)}{\sum_{i=1}^{n} Att_a(A_i)} \quad (3)$$

Where,
$Att_r(A_i)$ is the number of redefined pointcuts defined in aspect $A_i$,
$Att_a(A_i)$ is the number of available pointcuts defined in aspect $A_i$ and
$n$ is the number of aspects in the version of *AO* software.

4. *Aspect Inheritance Factor* (*AIF*): Aspect Inheritance Factor is the fraction of the sum of number of concrete aspects over the total number of available aspects defined in a version of *AO* software.

$$AIF = TCA/TAA \quad (4)$$

where,
$TCA$ is the total number of concrete aspects,
$TAA$ is the total number of available aspects and

5. *Class Method Inheritance Factor* (*CMIF*): Class Method Inheritance Factor is the fraction of the sum of number of redefined methods in all classes over the sum of number of available methods in all classes defined in a version of *OO* software.

$$CMIF = \frac{\sum_{i=1}^{n} M_r(C_i)}{\sum_{i=1}^{n} M_a(C_i)} \quad (5)$$

where,
$M_r(C_i)$ is the number of redefined methods defined in class $C_i$,
$M_a(C_i)$ is the number of available methods defined in class $C_i$ and
$n$ is the number of classes in the version of *OO* software.

6. *Class Inheritance Factor (CIF)*: Class Inheritance Factor is the fraction of the total number of extended classes to the total number of available classes defined in a version of OO software.

$$CIF = TEC/TAC \quad (6)$$

Where,
$TEC$ is the total number of extended classes,
$TAC$ is the total number of available classes.

VIII. CASE STUDY - UNIVERSITY AUTOMATION SYSTEM

An *SOA* application, with many core and cross-cutting concerns, is required to understand the impact of inheritance in *AO* software. Hence, an application should be selected in such a way that, it can be reused over versions. This requirement is satisfied by the University Automation System *(UAS)* Application that automates the operations of a typical university, since it possesses many scattered and tangled concerns. Hence *UAS* is selected as the case study to understand the impact of inheritance in *AO* software. Process flow diagram for the case study *UAS* is shown in Fig. 4. *UAS AJ 1.0* consists of only core concerns. *UAS AJ 1.1* consists of two cross-cutting concerns such as logging and persistence. Logging aspect is encapsulated in the Login web service of both the student and staff, in order to record who has logged-in and the timing when the user logs into the system. Persistence aspect is encapsulated to the Register web service of both categories of the users, in order to store the provided details persistently in the database. In logging aspect, audit log and event log inherits the functionality of *record events()* from file log. In persistence aspect, SQL translation is inherited from persistence. Abstract persistence consists of database connection. All the insert and update operations in database are given in concrete aspects.

One more cross-cutting concern namely security is identified as the cross-cutting concern in the version *UAS AJ 1.2*. Since, a single login concern is scattered for different users, this authentication functionality can be separated as a security aspect in *AOP*.

*UAS AJ 1.3* consists of observer pattern as cross-cutting concern. Once the result has been updated, the student can be notified that the result has been updated. For this purpose, Observer pattern can be used. Register observer can act as the abstract aspect. Notify observer can act as the concrete aspect. *UAS AJ 1.4* consists of exception handling as cross-cutting concern. The implementation of Exception handling in traditional *OOP* results in tangling of the exception handling concern and the primary concern. This implicit coupling between concerns leads to maintenance and evolution problems. To avoid this, code to handle exceptions in *UAS* can be separated as aspect of *AOP* so that whenever exceptions are thrown in *UAS*, those exceptions can be effectively handled in this version of *UAS*. IO exception, class not found exception and runtime exception inherits the functionality of *displayexception()* from abstract exception handling aspect.

## IX. ASPECT ORIENTED SOFTWARE REUSABILTY MEASUREMENT TOOL

The design of the *AOSRM* tool consists of four different modules to integrate the process flow and to calculate the values of the proposed metrics. The architecture of *AOSRM* tool is shown in Fig. 5.

### A. Module 1: Main

This module provides an interface to the user for selecting the directory where the classes and aspects are stored for a version of *AO* software. The evaluator can choose one version at a time in order to evaluate the values of the proposed metrics.

### B. Module 2: File Identifier and Logger

This module uses the selected folder location from the previous module. The folder is opened to identify all the file names with *.java* and *.aj* extension. The absolute path of each identified file is written in a log file.

### C. MODULE 3: SIGNATURE EXTRACTOR AND LOGGER

This module opens the log file created in the previous module and reads the absolute path names in the file. The path name is used to open the file and read its content to extract the signatures listed below. The extracted signatures are written back at the end of the same log file.

- Signature of the classes.
- Signature of the redefined classes.
- Signature of the methods.
- Signature of the redefined advice.
- Signature of the pointcut.
- Signature of the redefined pointcut.
- Signature of the redefined methods.
- Signature of the advice.
- Signature of the redefined attribute.
- Signature of the aspects.

### D. Module 4: Metrics Calculator and Logger

This module takes the input from the previous module and counts the total number of aspect constructs such as available advices, redefined advices, available pointcuts, redefined pointcuts, available attributes, redefined attributes, available aspects, concrete aspects, available methods, redefined methods, available classes, redefined classes. These values are used to calculate the proposed metrics such as Advice Inheritance Factor, Attribute Inheritance Factor, Pointcut Inheritance Factor, Aspect Inheritance Factor, Class Method Inheritance Factor and Class Inheritance Factor. The calculated values of the metrics are written back at the end of the same log file.

## X. MEASUREMENTS AND DISCUSSION

Various Java and AspectJ versions of *UAS* are given as input to the *AOSRM* tool. The tool calculates the metric values for the different versions and these values are tabulated in Tables I and II. The value of inheritance metrics are measured and compared across versions in order to analyze the reusability of *UAS*.

The bar chart shown in Fig. 6 is a plot of the metric values for the Java versions of *UAS* and the bar chart shown in Fig. 7 is a plot of obtained metric values for *AJ* versions of *UAS*.

### A. Measurements between OO and AO

There are no aspects in all the Java versions of UAS and hence the values of relevant metrics to measure the constructs of aspects such as *AdIF*, *PIF and AIF* are 0. The levels of aspect inheritance hierarchy are increased over *UAS AJ* versions. In *UAS AJ 1.1*, there are two aspects namely logging and persistence. These aspects in the UAS *AJ* version were described in detail in Section VIII. In *UAS AJ 1.1*, Audit log inherits the file log. The values obtained for the *AdIF, PIF, AIF* in this version are 0.5, 0.25, and 0.5. In *UAS AJ 1.2*, event log also inherits file log and SQL translate persistence inherits the database persistence. This version of *UAS* models the security functionality using a separate aspect. These increases in inherited artifacts are measured using the proposed metrics *AdIF, PIF, AIF*. Hence, the values obtained for the same aspect metrics are 0.625, 0.333, and 0.625. In order to encapsulate the exception handling functionality which is scattered in *OO*, one more aspect is added in UAS *AJ 1.3*. In this version of *UAS*, aspect for database security extends the aspect encapsulating the login security. Hence, the values obtained for the same aspect metrics are 0.7, 0.411, and 0.666. In version *UAS AJ 1.4*, the functionality modeling the observer pattern is defined as an aspect. The exception handling aspect defined in *UAS AJ 1.3* has been extended in this version to another aspect, which can add class not found exception and run time exception functionalities. Thus, the values obtained for the same aspect metrics are 0.75, 0.473, and 0.692. Therefore, the depth of inheritance hierarchy of aspects has been increased over the *AJ* versions of *UAS*. This increase in inheritance level is reflected by the metric values proposed for the measurement of inheritance.

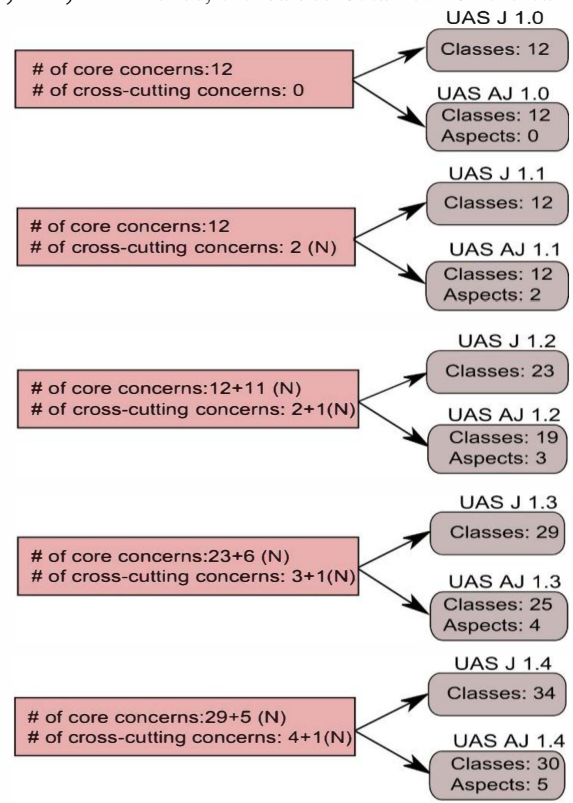

Fig 4. Applying *AO* across versions of *UAS*

Logging is one of the non-functional requirements implemented in the Java versions of *UAS*. Whenever a user logs in or logs out, the *log()* method is called to record the action for historical purposes. Since, this non-functional requirement is scattered across other functionalities, it is refactored and modeled as a logging aspect. This results in decrease in the values of *CMIF* for the *AJ* versions compared to the Java versions of *UAS*.

TABLE II. METRIC VALUES FOR ASPECTJ VERSIONS OF *UAS*

| S.No | Version   | AdIF  | PIF   | AttIF | AIF   | CMIF  | CIF   |
|------|-----------|-------|-------|-------|-------|-------|-------|
| 1    | UAS AJ 1.0 | 0.0   | 0.0   | 0.0   | 0.0   | 0.785 | 0.857 |
| 2    | UAS AJ 1.1 | 0.5   | 0.25  | 0.0   | 0.5   | 0.357 | 0.714 |
| 3    | UAS AJ 1.2 | 0.625 | 0.333 | 0.019 | 0.625 | 0.212 | 0.5   |
| 4    | UAS AJ 1.3 | 0.7   | 0.411 | 0.016 | 0.666 | 0.184 | 0.388 |
| 5    | UAS AJ 1.4 | 0.75  | 0.473 | 0.015 | 0.692 | 0.209 | 0.347 |

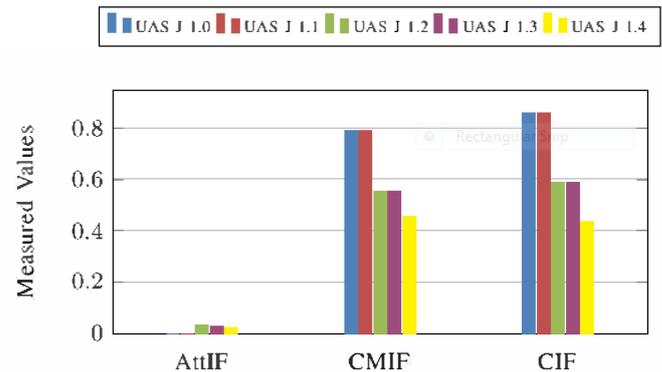

Fig. 6 Measured values for Java versions of UAS

There are several types of scattered login classes modeled in the Java versions of *UAS* responsible to check the authentication for users in different domains. Whereas, in AspectJ versions, only one login class is defined with a security aspect that is woven to the login class to authenticate the users. This reflects in the decrease of the values of *CIF* over the *AJ* versions compared to the Java versions.

B. *Effect of Inheritance*

Through inheritance, reusability has been achieved in *AO* software. Functionalities which occur very often in different modules of *OO* software have been encapsulated as cross-cutting concerns in *AO* software. Hence, the depth of inheritance in classes has been reduced over the AspectJ versions. Therefore, the complexity of the software is decreased and the modularity is increased. In each version of *AO* software, the depth of the inheritance hierarchy is increased in order to reuse the existing functionalities defined in the previous versions.

C. *Effect on reusability*

The multi-level inherited functionalities, such as logging, persistence and authentication have been scattered in Java versions of *UAS* which leads to duplication of reusable elements. Whereas, in aspectized versions the same functionalities have been neatly modularized achieving a better form of reusability. Since the level of inheritance applied in aspects was limited to a maximum of three in the design of *UAS AJ* versions, the resultant complexity is manageable and not high. From the obtained values of proposed metrics, the inheritance factor of classes, methods and attributes

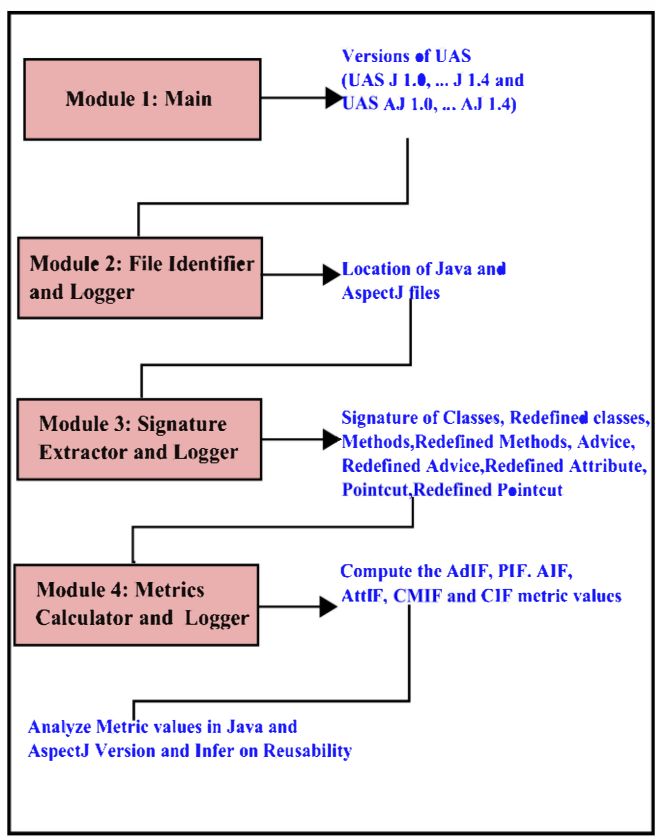

Fig 5. Architecture of *AOSRM* Tool

.TABLE I. METRIC VALUES FOR JAVA VERSIONS OF *UAS*

| S.No | Version   | AdIF | PIF | AttIF | AIF | CMIF  | CIF   |
|------|-----------|------|-----|-------|-----|-------|-------|
| 1    | UAS J 1.0 |      |     | 0.0   |     | 0.785 | 0.857 |
| 2    | UAS J 1.1 |      |     | 0.0   |     | 0.785 | 0.857 |
| 3    | UAS J 1.2 | NA   | NA  | 0.035 | NA  | 0.555 | 0.588 |
| 4    | UAS J 1.3 |      |     | 0.029 |     | 0.555 | 0.588 |
| 5    | UAS J 1.4 |      |     | 0.026 |     | 0.454 | 0.434 |

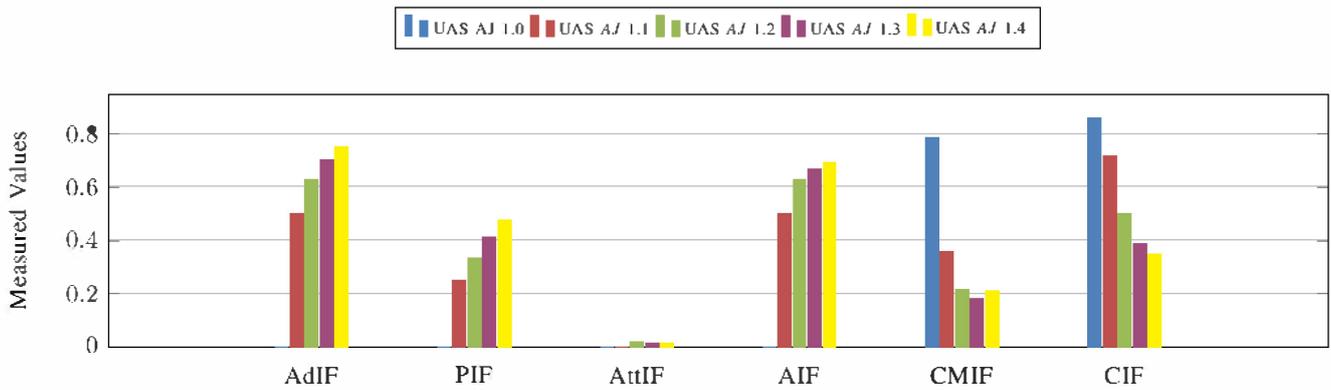

Fig. 7 Measured values for AJ versions of UAS

have decreased in *AO* versions and the inheritance factors of aspect, pointcut and advice have increased. Hence, inheritance is increased in *AO* elements which are based on higher reusable *AO* components.

## XI. CONCLUSIONS AND FUTURE WORK

Reusability has been considered as one of the important quality attributes signifying the efficient re-utilization of modeled artifacts. In this paper, an *AO* Reusability Evaluation Model has been proposed to infer on the effect of applying inheritance in *AOSD* methodology. The manifestations of inheritance in the artifacts of *AO* software have been defined and measured using a set of proposed metrics. In order to calculate the values of the proposed metrics, an automated *AOSRM* tool, has been designed, developed and tested. Five versions of *SOA* based *UAS* software was developed in both Java and AspectJ programming languages. The metrics were applied to all the five versions and measured using the *AOSRM* tool. Reusability of functionalities has been achieved through inheritance in all the versions of *UAS*. Based on the measured values, it was found that reusability of the elements of the aspects was found to be higher than that of the elements of classes. This work can be extended by applying the model to more number of case studies and obtain a general perspective of applying inheritance in *AO* software.